\documentclass[12pt,preprint]{aastex}

\begin{document}
 
\title{Study of the Cir X--1 Broad Band 
Spectrum at Orbital Phases Close to the Apoastron}

\author{R. Iaria\altaffilmark{1}, T. Di Salvo\altaffilmark{2},
N. R. Robba\altaffilmark{1}, L. Burderi\altaffilmark{3}}
\altaffiltext{1}{Dipartimento di Scienze Fisiche ed Astronomiche, 
Universit\`a di Palermo, Via Archirafi 36, 90123 Palermo, Italy;
iaria@gifco.fisica.unipa.it.}
\altaffiltext{2}{Astronomical Institute "Anton Pannekoek," University of 
Amsterdam and Center for High-Energy Astrophysics,
Kruislaan 403, NL 1098 SJ Amsterdam, the Netherlands.}
\altaffiltext{3}{Osservatorio Astronomico di Roma, Via Frascati 33, 
00040 Monteporzio Catone (Roma), Italy.}


\begin{abstract}
We report on the results of a BeppoSAX (1.8--200~keV) observation of
the peculiar X--ray binary source Circinus X--1 (Cir X--1) at the
orbital phases between 0.61 and 0.63.  We find that three components
are needed to fit the broad band spectrum: a blackbody component, at a
temperature of $\sim 0.6$ keV, a Comptonized component, with a seed-photon
temperature of $\sim 1.2$ keV, electron temperature of $\sim 6$ keV
and optical depth of $\sim 1.7$, and a power-law component dominating 
the spectrum at energies
higher than 20 keV.  We interpret the blackbody as the emission from
the accretion disk, while the Comptonized component probably comes
from a corona surrounding the inner part of the system.  This spectrum
is different from that observed at the periastron (Iaria et al. 2001a)
because of the presence of the blackbody component. We discuss the
implications of this difference and the presence of the power-law
component.
\end{abstract}

\keywords{accretion discs -- stars: individual: Cir~X--1 --- stars: 
neutron 
stars --- X-ray: stars --- X-ray: spectrum --- X-ray: general}

\section{Introduction}

Cir X--1 is a peculiar Low Mass X-ray binary (LMXB).  Generally in
LMXBs the companion star has a mass $\la 1$ M$_{\odot}$ and the orbit
is almost circular; contrarily Cir X--1 has a companion star of $\sim
3-5$ M$_{\odot}$ (probably a subgiant, see Johnston et al., 1999), an
orbital period of $\sim 16.6$ days and an eccentric orbit with $e\sim
0.7-0.9$ (Murdin et al., 1980; Tauris et al., 1999). It is a widely
accepted idea that Cir X--1 is a runaway binary, because of its
association with the supernova remnant G321.9-0.3 (Clark et al.,
1975). The most recent estimation of the distance to the supernova
remnant G321.9-0.3, and hence to Cir X--1, is 5.5 kpc (Case \&
Bhattacharya, 1998). Therefore in the following we adopt 5.5 kpc as
the distance to the source.  

The spectral and fast-timing properties of accreting low magnetic
field neutron stars permit distinguishing them into two classes, Z and
atoll sources, using the pattern each source describes in an X-ray
color-color diagram (Hasinger \& van der Klis 1989). The Z sources are
very bright and could have a moderate magnetic field, while the atoll
sources have a lower luminosity and probably a weaker magnetic
field. The timing analysis of RXTE data (Shirey
et al., 1996) evidenced that Cir X--1 is more similar to LMXBs of the
Z class.  However
no quasi periodic oscillations at kilohertz frequencies were observed
in Cir X--1 (Shirey, 1996) contrarily to all the other Z-sources.

Brandt et al. (1996), using ASCA/GIS data taken on 1994 August 4--5,
found a sudden variation (with a timescale of $\sim 20$ minutes) of the
flux at the periastron phase, where the count rate increased from
$\sim 30$ counts s$^{-1}$ to $\sim 300$ counts s$^{-1}$. In the
low-count rate state they obtained a good fit to the 0.6--10 keV
spectrum using a partial covering component, in addition to a
two-blackbody model, with a corresponding hydrogen column of $\sim
10^{24}$ cm$^{-2}$; in the high-count rate state the partial covering
of the X-ray spectrum was not needed. Because of the similarity with
the spectra of Seyfert 2 galaxies with Compton-thin tori, Brandt et
al.  (1996) suggested that matter at the outer edge of the accretion
disk, together with an edge-on disc orientation, could explain the
partial covering of the spectrum in Cir X--1.

Iaria et al. (2001a), using BeppoSAX data, analysed the 0.1--100 keV spectrum 
of Cir X--1 at orbital phases near the periastron. They found that the
continuum could be well described by either a two-blackbody model or
a Comptonization model.  In the first case, the so-called
``Eastern Model'' proposed by Mitsuda et al. (1984), the soft
blackbody emission is interpreted as the emission from the inner
region of the accretion disk while the hard blackbody comes from the
neutron star surface.  In the case of Cir X--1 the radius of the soft
blackbody was $\sim 140$ km, which should correspond to
the inner radius of the accretion disk, while the radius of the harder
blackbody component was $\sim 30$ km, that is too large to be interpreted
as the neutron star radius. Therefore the Comptonization model was
preferred.  A hard power-law tail, dominanting the spectrum at
energies higher than 15 keV, was also observed.  A similar feature was
observed in other Z-source LMXBs (GX 17+2, Di Salvo et al., 2000; GX
349+2, Di Salvo et al., 2001; GX 5--1, Asai et al., 1994;  Sco X--1,
D'Amico et al., 2000), suggesting that Cir X--1 shows a behaviour
similar to that of other Z-sources. Finally a strong absorption edge
at $\sim 8.4-8.7$ keV, produced by highly ionized iron, was needed to
fit the spectrum of Cir X--1.

Brandt \& Schulz (2000), using Chandra data, observed Cir X--1 near the
periastron. They found, for the first time in the X-ray band,
P-Cygni profiles of lines emitted by highly ionized matter. The
outflow velocity necessary to explain the P-Cygni profiles was $\sim
2000$ km/s.  Under the hypothesis that the source is seen nearly
edge-on, they supposed that the model of a thermally driven wind
(Begelman et al., 1983) could explain this outflow of ionized matter
from the system. Also they suggested that the absorption edge
observed by Iaria et al. (2001a) could be connected to the outflow.
   
Iaria et al. (2001b), using ASCA data, studied the energy spectrum of
Cir X--1 along its orbit. They distinguished three different X-ray
states of the source as a function of its orbital phase. At first,
when the source is near the periastron (according to the ephemeris of
Stewart et al., 1991), the luminosity is super-Eddington and a large
X-ray flaring activity is present. In the second state, corresponding
to the orbital phase interval $0.2-0.6$, the luminosity is
sub-Eddington and the lightcurve does not show flaring activity. In
the third state, in the orbital phase interval $0.7-0.9$, an excess of
neutral matter (probably the atmosphere of the companion star and/or
the accreting matter) partially covers the emitting region.  In all
these states, the continuum was well fitted using a Comptonization
model.  Unfortunately, because of the relatively narrow band of ASCA
(0.6--10 keV), the spectral evolution of the hard tail along the orbit
could not be studied. On the other hand, it was possible to study the
evolution of the absorption edge along the orbit of Cir X--1. At the
phase of the flaring activity the hydrogen column derived from the
edge was $\sim 10^{24}$ cm$^{-2}$; when the source moved away from the
periastron the hydrogen column decreased; the absorption edge (from
ionized iron) was no longer detected at orbital phases from 0.78 to the
periastron.  It is possible that the outflowing wind is connected to
the huge flaring activity at the periastron.

In this work we present the broad band (1.8--100 keV) spectrum of Cir
X--1 from a BeppoSAX observation taken at  orbital phases
0.61--0.63. The broad band allows us to constrain the continuum
components of the spectrum and to detect the presence of a hard tail at
energies higher than 20 keV.

\section{Spectral Analysis of Cir X--1 at the Apoastron}

A pointed observation of Cir X--1 was carried out between 1999 Feb 07
03:15:19 UT and 1999 Feb 08 19:25:33 UT, with the Narrow Field
Instruments, NFIs, on board BeppoSAX (Boella et al. 1997).  These
consist of four co-aligned instruments covering the 0.1--200~keV
energy range: a Low-Energy Concentration Spectrometer (LECS; operating
in the range 0.1--10~keV), two Medium-Energy Concentration
Spectrometers (MECS; 1.3--10~keV), a High-Pressure Gas Scintillation
Proportional Counter (HPGSPC; 7--60~keV), and a Phoswich Detector
System (PDS; 13--200~keV).  The LECS and the HPGSPC were not active
during this observation.  The effective exposure time was 25~ks in the
MECS and 11~ks in the PDS.  In the MECS image, data were extracted
from a circular region of $4'$ radius centered on the source centroid.
Data extracted from the same detector region during blank field
observations were used for background subtraction.  Background
subtraction for the PDS data was obtained from data accumulated during
off-source intervals. We rebinned the energy spectra in order to have
at least 25 counts/channel and to oversample the instrument resolution
with the same number of channels at all energies\footnote{see the
BeppoSAX cookbook at http://www.sdc.asi.it/software/index.html}.

Using the orbital parameters reported by Stewart et al. (1991), we
have calculated that this observation corresponds to the orbital phase
interval 0.61--0.63, just after the apoastron.  In Figure 1 we plotted the
hardness-intensity diagram, where the hardness is the
ratio between the count rate in the 3--7 keV energy band to the count
rate in the 1--3 keV energy band, and the intensity is the count rate
in the 1--7 keV energy band. In this figure we show both the data of
this observation and those of a previous BeppoSAX observation at the
periastron (see Iaria et al., 2001a). The data at the apoastron show
larger hardness and lower intensity with respect to the data at the
periastron.  In Figure 2 we plotted the color-color diagram of
Cir X--1 data, where the hard color is the ratio between the count
rate in the energy band 7--10 keV to that in the energy band 3--7 keV
and the soft color is the hardness defined above.  The data at
the apoastron are at the top right in this figure while the data
obtained at the periastron are at the bottom left, implying that the
spectrum at the apoastron is harder in both the soft and hard color.
In Figure 3 we plotted the lightcurve of the data at the apoastron in
the energy band 1.8--10 keV (upper panel) 
together with the soft color as a function of time
(lower panel).  The count rate smoothly decreases from $\sim 300$
Counts s$^{-1}$, at the beginning of the observation, to $\sim 280$
Counts s$^{-1}$, at the end of the observation. No flaring activity
is present.  

Because no significant spectral variation is observed in the hardness ratios
described above, we produced an unique spectrum, in the band 1.8--200
keV, from this observation.  A systematic error of 1\% was added to
each spectrum.  As customary, in the spectral fitting procedure we
allowed for a different normalization of the PDS spectrum relative to
the MECS spectrum, always checking that derived values
are in the standard range for that instrument\footnote{see the
BeppoSAX cookbook at http://www.sdc.asi.it/software/index.html}.  
The energy ranges used for the spectral analysis are 1.8--10 keV for the 
MECS and 15--200 keV for the PDS.  
Because of the lack of low energy data we fixed the value
of the average photoelectric absorption by cold matter at $ 1.7
\times 10^{22}$ cm$^{-2}$ (see e.g.  Iaria et al. 2001a, 2001b).
We started fitting the data to the Comptonization model {\it Comptt}
(Titarchuk, 1994). This model gave an unacceptable fit with
$\chi^2/d.o.f.=1161/100$; the residuals with respect to this model are
shown in Figure 4 (upper panel). From the residuals, it is evident that there 
is a feature between 8 and 10 keV and an excess at
energies higher than 15 keV.  Note that because this simple model does
not fit the high energy part of the spectrum, we had to fix the PDS
normalization factor to 0.85 (its typical value) to avoid unreasonably 
high values of this parameter.

Then, in agreement with the model proposed by Iaria et al. (2001a) for the
spectrum of Cir X--1 at the periastron, we added a power law to fit the
high energy excess, an absorption edge at $\sim 8.4$ keV and a
Gaussian emission line at $\sim 6.7$ keV (with the Gaussian $\sigma$
fixed at 0.25 keV), keeping fixed the PDS normalization factor. The
addition of each of these components is statistically significant and 
in this way the fit improves significantly, giving $\chi^2/d.o.f.=117/94$.  
In Table 1 (Model 1) we present the best fit parameters obtained for 
this model.  Looking at the residuals with respect to Model 1 (Figure 4, 
middle panel), we still see modulations between 1.8 and 5
keV and between 20 and 40 keV, indicating that the continuum emission
could be more complex.

The modulation in the residuals could be due to the large contribution at low 
energies of the power law which has a large normalization and a photon 
index of $\sim 3.3$. We modified the power law multiplying it for an
exponential cutoff at low energy ({\it Expfac} in XSPEC), fixing the
low-energy cutoff at the electron temperature of the Comptonized component. 
In this way we found a stable (and reasonable) fit, with the PDS normalization 
factor and the width of the Gaussian emission line as free parameters (see
Table 1, Model 2). The residuals (see Fig. 4, lower panel) show that
the modulation between 1.8 and 5 keV is now smaller.  We obtain
$\chi^2/d.o.f.=96/92$ and an F-Test with respect to the previous
model gives a probability of chance improvement of the fit of $\sim
1.1 \times 10^{-4}$. However, the modulation between 20 and 40 keV is
still visible.

We, therefore, tried an alternative model, consisting of Model 1 to
which we added a blackbody component at low energies.
The addition of the blackbody component improved the fit
significantly, giving $\chi^2/d.o.f.=71/91$, with a probability of
chance improvement with respect to  Model 1 of $\sim 6.5 \times
10^{-10}$.  Hence we consider this as the best fit model.  The F-Test for 
the addition of the power law component in this case gives a probability 
of chance improvement of $\sim 9.8 \times 10^{-10}$.
In Table 1 (Model 3) we report the parameters of the best fit model.
The blackbody temperature is $kT_{bb} \sim 0.59$ keV.  The parameters of the
Comptonized component are as follows: the seed-photon temperature is 
$kT_{0} \sim 1.2$ keV, the electron temperature is $kT_{e} \sim 6$ keV and 
the optical depth of the Comptonizing cloud is $\tau \sim 1.7$.  
The centroid of the Gaussian emission
line is at $\sim 6.8$ keV, with FWHM$\sim 0.39$ keV, and the
absorption edge is at $\sim 8.47$ keV, with an optical depth
$\tau_{edge} \sim 0.07$. The PDS normalization factor is $\sim 1$.  

In Figure 5 we show the data and the residuals with respect to the Model
3.  Note that in this case the modulations between 1.8 and 5 keV and
between 20 and 40 keV disappear, confirming that Model 3 gives the
best fit to these data.  In Figure 6 we show the corresponding
unfolded spectrum.
The extrapolated unabsorbed flux, in the energy band 0.1--200 keV, is
$\sim 2.3 \times 10^{-8}$ erg s$^{-1}$ cm$^{-2}$ for the Comptonized
component, $1.4 \times 10^{-8}$ erg s$^{-1}$ cm$^{-2}$ for the
blackbody component and $1.35 \times 10^{-8}$ erg s$^{-1}$ cm$^{-2}$
for the power-law component.

\section{Discussion}

We analyzed data of Cir X--1 from a BeppoSAX observation (in the
orbital phase range 0.61--0.63) in the energy range 1.8--200 keV.  The
lightcurve and the hardness ratio do not show large variations during
the observation.  From Figures 1 and 2 it is clear that the spectrum,
in the range 1--10 keV, is different from that observed at the
periastron (Iaria et al., 2001a); in fact the spectrum at the
apoastron is harder in both the soft and hard colors and the intensity
is lower than that at the periastron.  
The spectrum of Cir X--1 during this observation
is very similar to that of  other Z-sources
(see Di Salvo et al., 2000; Di Salvo et al., 2001). The soft X-ray
continuum in the energy range 1.8--10 keV is well fitted by a blackbody
and a Comptonized component, while a power-law component, with photon
index $\sim 3$, is present at energies higher than 20 keV.

The blackbody temperature is $\sim 0.6$ keV and the corresponding
radius of the emitting region is $\sim 58$ km.  This radius is too
large to be associated with the emission from the neutron star surface;
therefore the blackbody component is probably emitted by the inner
part of the accretion disk.  The seed-photon and electron temperature
of the Comptonized component are $\sim 1.2$ keV and $\sim 6$ keV,
respectively, and the optical depth is $\sim 2$. We calculated the
radius, $R_W$, of the seed-photon emitting region using the parameters
reported in Table 1 (Model 3) following in 't Zand et {\it al.}
(1999); we obtain $R_W \sim 17$ km.  This component could be produced
by a corona surrounding the neutron star and occulting the inner
region of the accretion disk.
 
We found an absorption edge at $\sim 8.5$ keV with $\tau_{edge} \sim
0.07$.  The energy of the absorption edge is compatible with iron
ionization levels of Fe XXIII--XXIV (see Turner et {\it al.}, 1992 for
a correspondence between iron edge energy and ionization level).  The
best fit value for the optical depth $\tau_{edge}$, considering the
photoionization cross section for the K-shell of Fe XXIII (Krolik \&
Kallman, 1987), corresponds to a hydrogen column density of $ \sim 1.2
\times 10^{23}$ cm$^{-2}$.  A Gaussian emission line is also detected 
at $\sim 6.8$ keV.  The energy of the emission line corresponds to 
iron ionization levels Fe XXIV--XXV, compatible with that of the
absorption edge (see Turner et al., 1992).  This suggests that the
iron emission  line could be produced in the same region where the
absorption edge is produced.  Note that the ASCA spectrum at orbital
phase 0.54--0.56, the nearest to the orbital phase of the BeppoSAX 
observation analyzed here, shows both the iron emission line and the 
absorption edge, and their parameters are in agreement with the values
we found from this BeppoSAX observation (cf. Iaria et al. 2001b).

The best fit model obtained here is different from that used by 
Iaria et al. (2001a) for the BeppoSAX spectrum of Cir X--1 taken at the 
periastron, when a flaring activity was present.  In that case only the 
Comptonized component was needed to fit the continuum from 0.12 to 10 keV 
and the blackbody component was not needed. The seed-photon  and 
electron temperatures at the periastron were $\sim 0.4$ and $\sim 1$ keV, 
respectively, with a radius of the seed-photon emitting region  of 
$ \sim 140$ km.

One possibility to explain the large seed-photon radius observed at the 
periastron, in the hypothesis it is the inner radius of the accretion disk, 
might be the presence of a large neutron star magnetic field with
a strength of $\sim 4 \times 10^{10}$ G, more the one order of
magnitude higher than the typical value estimated for the
Z-sources. 
In light of the new results on Cir X--1, presented in this
work, we can exclude this possibility.
In fact, the flux from the source was larger than $5.4 \times 10^{-8}$ erg 
cm$^{-2}$ s$^{-1}$ during the observation at the periastron and, for a
distance to the source of 5.5 kpc (Case \& Bhattacharya, 1998), the
corresponding luminosity is super-Eddington.  During the observation
analysed here, where Cir X--1 is near the apoastron, the estimated
flux is $3.8 \times 10^{-8}$ erg cm$^{-2}$ s$^{-1}$ and the luminosity
is sub-Eddington.  Because the X-ray luminosity is connected to the
accretion rate, the ram pressure of the disk should be lower, by a factor 
of $\sim 1.4$ at the apoastron than at the periastron. Consequently the 
magnetospheric radius (see e.g. Burderi et al., 1998) should be 
$\sim 154$ km at the apoastron, assuming the same strength of the magnetic 
field of $\sim 4 \times 10^{10}$ G. However we find that the inner radius 
of the accretion disk is $\sim 58$ km.

Another possibility to explain the large inner radius of the accretion 
disk observed at the periastron can be the occultation of the inner region 
by a corona surrounding the neutron star.  This corona should be therefore 
more extended at the periastron and cover a larger part of the accretion
disk.  This hypothesis is also supported by the differences between the 
spectrum at the periastron and at the apoastron.  The seed-photon 
temperature is much lower at the periastron ($\sim 0.4$ keV), with a 
comparatively higher accretion rate, than at the apoastron ($\sim 1.2$ keV), 
implying that at the periastron the seed photons for the 
Comptonization come from a further region in the accretion disk.
In particular at the periastron the seed photons probably come from a 
region which extends up to a radius of $\sim 140$ km in the accretion disk
(see Iaria et al. 2001a).
On the other hand at the apoastron, the seed photons probably come from
a region which extends up to $\sim 20$ km in the accretion disk.
This also explains why we observe a blackbody component at the apoastron,
which we interpret as the emission of the part of the accretion disk not
covered by the corona, i.e.\ from radii larger than $\sim 60$ km, while
this component is not observed at the periastron.  In fact the disk emission 
from radii larger than 150 km would probably be too soft to be 
significantly detected by the BeppoSAX/LECS.

A power-law component, dominanting the spectrum of Cir X--1 at energies 
higher than 20 keV, is also detected with high statistical significance. 
This component was observed two times in Cir X--1:
at the periastron (Iaria et al., 2001a), and near the apoastron
(present work).  At the periastron of Cir X--1 the photon index was
$\sim 3.3$ with a flux in the band 10--200 keV of $\sim 1.4 \times
10^{-10}$ erg s$^{-1}$ cm$^{-2}$. At the apoastron the photon-index
is $\sim 2.9$ with a flux in the band 10--200 keV of  $\sim 1.1 \times
10^{-10}$ erg s$^{-1}$ cm$^{-2}$.  Note that the radio emission in 
Cir X--1 is maximum near the periastron (Stewart et al., 1993; Fender, 1997).
Also, the observation of Cir X--1 presented in
this work is at the orbital phase between 0.61--0.63; at the same orbital
phases Fender (1997) detected a secondary radio
flaring activity with a maximum intensity of $\sim 25$ mJy, lower than
at the periastron (where the maximum intensity is $\sim 40$ mJy).
Although a systematic study of the presence of the hard power-law component 
along  the entire orbit is needed to draw a conclusion, 
the coincidence between 
the presence of this component and periods of radio flares suggests a 
connection between these two phenomena. 
In particular the power-law component might have a non-thermal
origin and might be related to the presence of the high velocity electrons
responsible for the radio emission.

As mentioned above,
there is evidence that a jet is present at the periastron 
(Stewart et al., 1993; Fender, 1997). Also, 
the prominent absorption edge from highly ionized matter observed
at the periastron of Cir X--1 is probably connected with the outflowing 
wind observed by Chandra (Brandt \& Schulz, 2000; Iaria et al., 2001b).
These observations could be explained using a model similar
to that proposed for young stellar objects (YSOs). The source HH30
presents both a jet and a wind from the disk.  Goodson et al. (1999)
explain the presence of these two phenomenaby postulating that the
interaction between the stellar magnetosphere and the surrounding
accretion disk drives a two-component outflow, the jet and the disk
wind. The magnetic loops connecting the accretion disk to the star,
expand due to the differential rotation between the star and the disk
(Hayashi at al., 1996; Lovelace et al. 1995). The plasma trapped in
the loops is driven toward the rotation axis, giving a collimated jet
and an outflowing wind outward along the disk surface.
Magnetic reconnections can produce large X-ray flaring activity and
allow for the process to repeat.

In this interpretation the P-Cygni profiles and the absorption edge
observed near the periastron could be explained by the disk wind
expected in this model, the radio emission by the collimated jet and
the flaring activity  at the periastron by the magnetic
reconnections. The Comptonized emission observed at the periastron
could also come from the outflowing wind. The outflow and then the
Comptonizing region, because of the large magnetic activity, probably extends
far from the neutron star above the disk, and, as mentioned above, this might 
explain why we do not observe the blackbody emission from the accretion disk 
at the periastron. 
However, note that the work of Goodson et al. (1999) is based on YSOs and
possibly the results might not be applied in a straightforward way to
the case of Cir X--1, because of the different strength of the neutron
star magnetic field.  A study of the consequences of this model in the
more extreme case of magnetized neutron stars would be interesting in
this context.
Following this scenario we can suppose that at the apoastron there are no
magnetic reconnections and consequently the source 
is not highly variable.  However, the reason for this difference between
the periastron and the apoastron is not clear, although we can suppose it 
can be related to the difference in the accretion rate and/or tidal 
interactions, and should be investigated with a proper model.

\section{Conclusions}

In this work we presented the analysis of the broad band (1.8--200
keV) spectrum of Cir X--1 at the orbital phase 0.61--0.63, using a
BeppoSAX observation. The luminosity of the source is sub-Eddington, and
the spectrum is harder than that at the periastron (Iaria et al., 2001a).
The model used to fit this spectrum is the same that has been used
for other Z-sources (see Di Salvo et al., 2000; Di Salvo et al., 2001); 
we find that the continuum is well fitted using three
components. The first is a blackbody with a temperature of $\sim 0.6$
keV, probably produced by the inner region of the accretion disk. The
corresponding blackbody radius is $\sim 58$ km. The second is a
Comptonized component with a seed-photon temperature of $\sim 1.2$
keV, an electron temperature of $\sim 6$ keV and optical depth of
$\sim 1.7$. We interpret this component as the emission from a corona
(or an outflowing wind) around the neutron star. In this case there is
a weaker outflow of matter, suggested by the relatively low optical
depth of the absorption edge, which is now $\tau_{edge}\sim 0.07$, one
order of magnitude lower than at the periastron.  The third component
is a power-law that could be produced by Comptonization off
non-thermal electrons or by synchrotron emission in a jet. An emission
line is also present at $\sim 6.8$ keV, with $FWHM \sim 0.39$ keV.
The ionization level inferred from the line is Fe XXIV--XXV. The last
observed feature is an absorption edge from highly ionized iron (Fe
XXIII--XXIV), probably produced by an outflowing wind along the disk
surface.

\acknowledgments This work was partially supported by the Italian
Space Agency (ASI) and the Ministero della Ricerca Scientifica e
Tecnologica (MURST).

\clearpage

\section*{TABLE}
\begin{table}[th]
\begin{center}
\footnotesize 
\caption{Results of the fit of Cir X--1 1.8--200 keV spectrum at the binary 
phase 0.61--0.63. The best fit (Model 3) is
given by a blackbody, a Comptonized spectrum modeled by
Comptt, a power law, an absorption edge and a Gaussian emission line.
Uncertainties are at 90\% confidence level for a single
parameter. $kT_{\rm bb}$ and N$_{\rm bb}$ are, respectively, the blackbody
temperature and normalization in units of $L_{39}/D_{10}^2$, where $L_{39}$ 
is the luminosity in units of $10^{39}$ ergs/s and $D_{10}$ is the distance
in units of 10 kpc. $kT_0$ is the temperature of the seed photons for
the Comptonization, $kT_e$ is the electron temperature, $\tau$ is the
optical depth of the scattering cloud using a spherical geometry, 
N$_{\rm comp}$ is the normalization of the Comptt model in XSPEC v.11 units,
and f$_{bol}$ is the bolometric (unabsorbed) flux of the Comptonization 
component.  $EQW_{\rm Fe}$ indicates the equivalent
width of the Gaussian emission line, E$_{Fe}$ its centroid energy and
I$_{Fe}$ its intensity. The
power law normalization, N$_{\rm PL}$, is in units of photons
keV$^{-1}$ cm$^{-2}$ s$^{-1}$ at 1 keV.}

\begin{tabular}{l|c|c|c}

\tableline  
\tableline

 & Model 1 & Model 2 & Model 3 \\
& Comptt+  Power-law + &  Comptt+ Modified Power-law+  & BB+Comptt+ \\
& Line +Edge &  Line +Edge &  Power-law+ Line +Edge  \\        
                 
\tableline                               
  &  &  &    \\
$N_{\rm H}$ $\rm (\times 10^{22}\;cm^{-2})$ 
                       &  1.7 (fixed) &  1.7 (fixed) &  1.7 (fixed) \\

 E$_{edge}$ (keV)      & $8.40^{+0.13}_{-0.11}$  & $8.41^{+0.21}_{-0.15}$   
                       & $8.47^{+0.29}_{-0.24}$ \\

$\tau_{\rm edge}$     & $0.175 \pm 0.025$ & $0.125^{+0.023}_{-0.047}$
                      & $0.071^{+0.016}_{-0.035}$\\

$k T_{\rm bb}$ (keV) &--  & -- &  $0.589^{+0.036}_{-0.021} $ \\

 N$_{\rm bb}$        &--  & --&  $0.1710^{+0.0067}_{-0.0098} $ \\

$k T_0$ (keV)        &  $0.637^{+0.027}_{-0.023}$ & $0.569^{+0.037}_{-0.047}$
                     &  $1.182^{+0.042}_{-0.030}$  \\

$k T_{\rm e}$ (keV)   &  $1.641 \pm 0.014$ &   $1.510^{+0.027}_{-0.046}$ 
                      & $6.0^{+11.1}_{-2.6}$ \\

$\tau$                & $16.38^{+0.31}_{-0.53}$  & $19.14^{+0.86}_{-0.69}$
                      &  $1.72^{+1.96}_{-0.33}$\\

 N$_{\rm comp}$       &  $5.71^{+0.18}_{-0.15}$ &  $6.37^{+0.64}_{-0.28} $ 
                      &  $0.75^{+1.59}_{-0.23} $ \\

f$_{bol}$($\times 10^{-8}$ ergs cm$^{-2}$ s$^{-1}$)  & 3.2 & 3.2 & 2.3\\

E$_{Fe}$ (keV)        & $6.702^{+0.084}_{-0.100}$ & $6.757^{+0.087}_{-0.097}$ 
                      & $6.800^{+0.094}_{-0.100}$   \\

$\sigma_{\rm Fe}$ (keV)   & 0.25 (fixed) & $< 0.16$  & $< 0.15$ \\

I$_{\rm Fe}$ $(\times 10^{-2}$ ph cm$^{-2}$ s$^{-1}$)  
				 & $0.92^{+0.20}_{-0.23}$ 
                                 & $0.73^{+0.36}_{-0.22}$ 
                                 & $0.69^{+0.29}_{-0.24}$\\

$EQW_{\rm Fe}$ (eV)              & $47.7^{+10.1}_{-12.1}$ 
                                 & $39.6^{+19.9}_{-11.7}$  
                                 & $42.2^{+17.9}_{-13.3}$ \\

Photon Index                & $3.318^{+0.020}_{-0.066}$ 
                            & $3.593^{+0.065}_{-0.146}$
                            & $2.96^{+0.30}_{-0.53}$ \\

 N$_{\rm PL}$              & $6.15^{+0.57}_{-0.93}$ 
                            & $14.1^{+3.9}_{-5.9}$
                            & $0.89^{+1.92}_{-0.77}$  \\
Low-energy cutoff (keV)  &-- & $1.510^{+0.027}_{-0.046}$ & --\\
PDS normalization           & 0.85 (fixed) & $1.049^{+0.156}_{-0.033}$
                            & $0.98^{+0.15}_{-0.10}$\\

$\chi^2$/d.o.f.             & $117/94$ & $96/92$  & $71/92$ \\

\tableline
\end{tabular}
\end{center}
\end{table}

 \clearpage

\section*{FIGURE CAPTIONS}
\bigskip
\noindent
{\bf Figure 1}: Hardness-intensity diagram of Cir X--1.  We show in
the same box the data at the periastron binary phase (Iaria et al.,
2001a) and the new data taken near the apoastron phase, as indicated in the
figure. The hardness is defined by the  ratio [3-7 keV]/[1-3 keV] and  the
intensity of the source is calculated in the energy band [1-7 keV]. 
The bin time is 137 s.  
The data at the apoastron correspond to lower intensity and higher 
hardness (top left in the diagram). \\
{\bf Figure 2}: Color-color diagram of Cir X--1. We report here the
same data shown in Figure 1. The hard color is  the hardness ratio
[7-10 keV]/[3-7 keV] and the soft color is the hardness ratio [3-7
keV]/[1-3 keV].  The bin time is 137 s. The data at the apoastron are
at the top right of the box showing that the spectrum is harder in both
the soft and the hard color. \\
{\bf Figure3}: Cir X--1 lightcurve in the energy band 1.8--10 keV (MECS data, 
upper panel) and ratio of the count rate in the energy band 3--7 keV to that 
in 1.8--3 keV (lower panel).  The bin time is 137 s.\\
{\bf Figure 4}: Residuals in units of $\sigma$ with respect to the
Comptonization model (upper panel), to Model 1 (middle panel), to Model 2
(lower panel).  The models are described in the text and in Table 1. \\
{\bf Figure 5}: Energy spectrum (1.8--200 keV) of Cir X--1 at the
orbital phase 0.61--0.63.  Data and the corresponding best fit model
(see Tab. 1, Model 3) are shown in the upper panel, residuals in units
of $\sigma$ with respect to the best fit model are shown in the lower
panel.\\
{\bf Figure 6}: Unfolded spectrum of Cir X--1 and the best fit model,
shown in this figure as a solid line. The single components of the
model are also shown, namely the blackbody (dashed line), {\it Comptt}
(dotted line), power-law (dot-dot-dashed line) and iron emission line
(dot-dashed line).  The absorption edge at 8.45 keV is visible.

\clearpage

\begin{figure}
\plotone{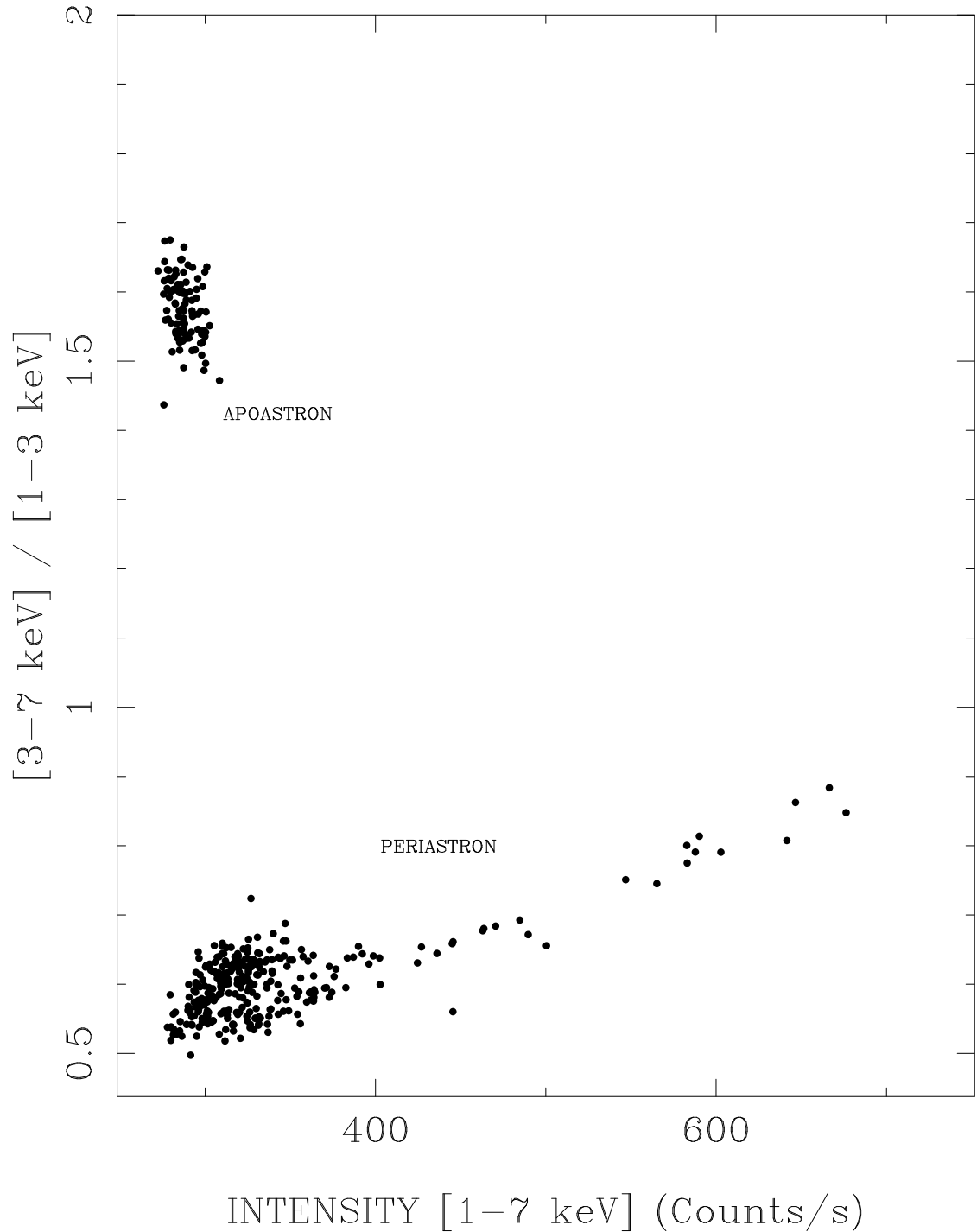}
\caption{\label{fig:fig1} Hardness-intensity diagram of Cir X--1.  We show in
the same box the data at the periastron binary phase (Iaria et al.,
2001a) and the new data taken near the apoastron phase, as indicated in the
figure. The hardness is defined by the  ratio [3-7 keV]/[1-3 keV] and  the
intensity of the source is calculated in the energy band [1-7 keV]. 
The bin time is 137 s.  
The data at the apoastron correspond to lower intensity and higher 
hardness (top left in the diagram).}
\end{figure}

\begin{figure}
\plotone{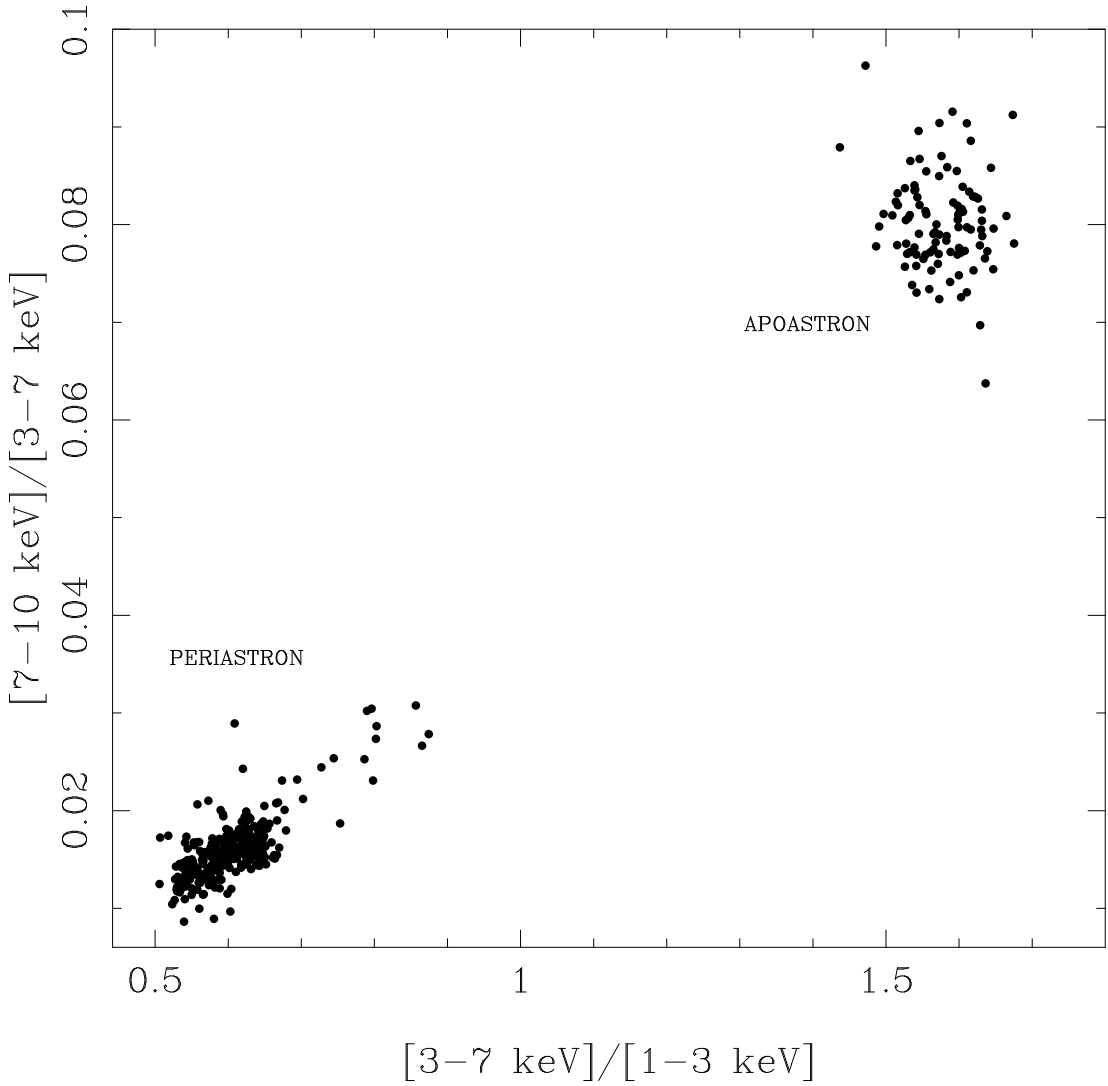}
\caption{\label{fig:fig2} Color-color diagram of Cir X--1. We report here the
same data shown in Figure 1. The hard color is  the hardness ratio
[7-10 keV]/[3-7 keV] and the soft color is the hardness ratio [3-7
keV]/[1-3 keV].  The bin time is 137 s. The data at the apoastron are
at the top right of the box showing that the spectrum is harder in both
the soft and the hard color. }
\end{figure}

\begin{figure}
\plotone{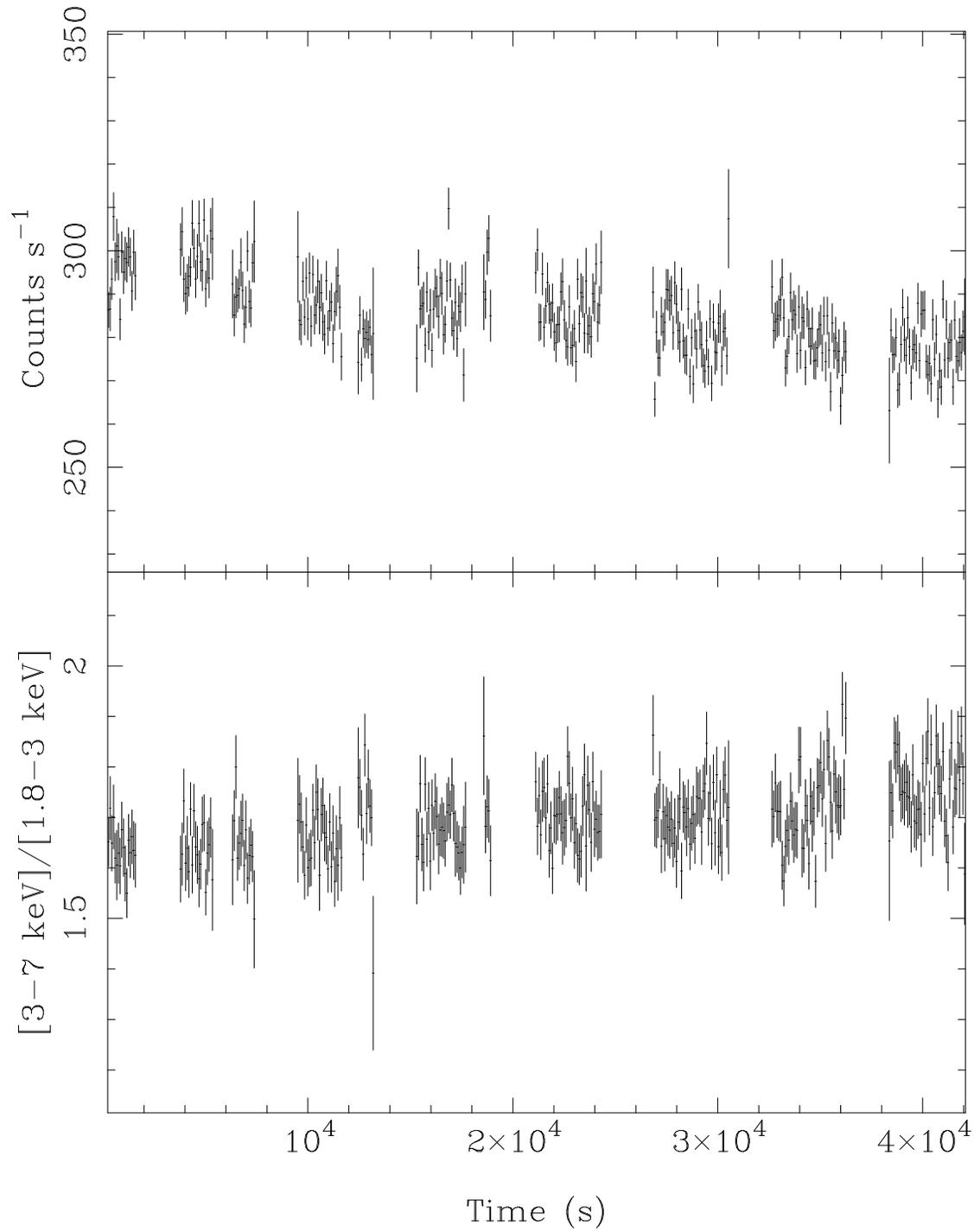}
\caption{\label{fig:fig3} 
Cir X--1 lightcurve in the energy band 1.8--10 keV (MECS data, 
upper panel) and ratio of the count rate in the energy band 3--7 keV to that 
in 1.8--3 keV (lower panel).  The bin time is 137 s.}
\end{figure}

\begin{figure}
\plotone{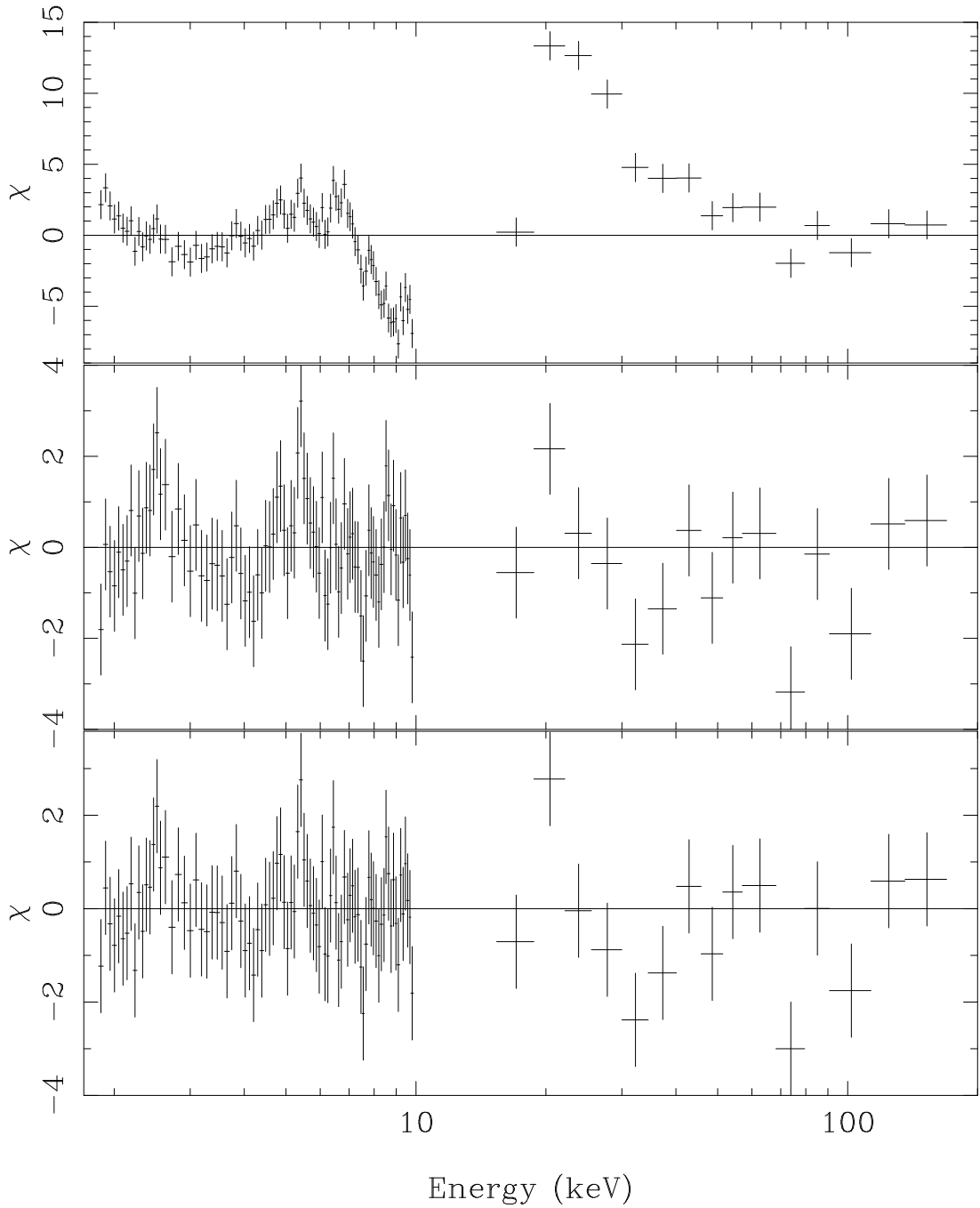}
\caption{\label{fig:fig4} Residuals in units of $\sigma$ with respect to the
Comptonization model (upper panel), to Model 1 (middle panel), to Model 2
(lower panel).  The models are described in the text and in Table 1. }
\end{figure}

\begin{figure}
\plotone{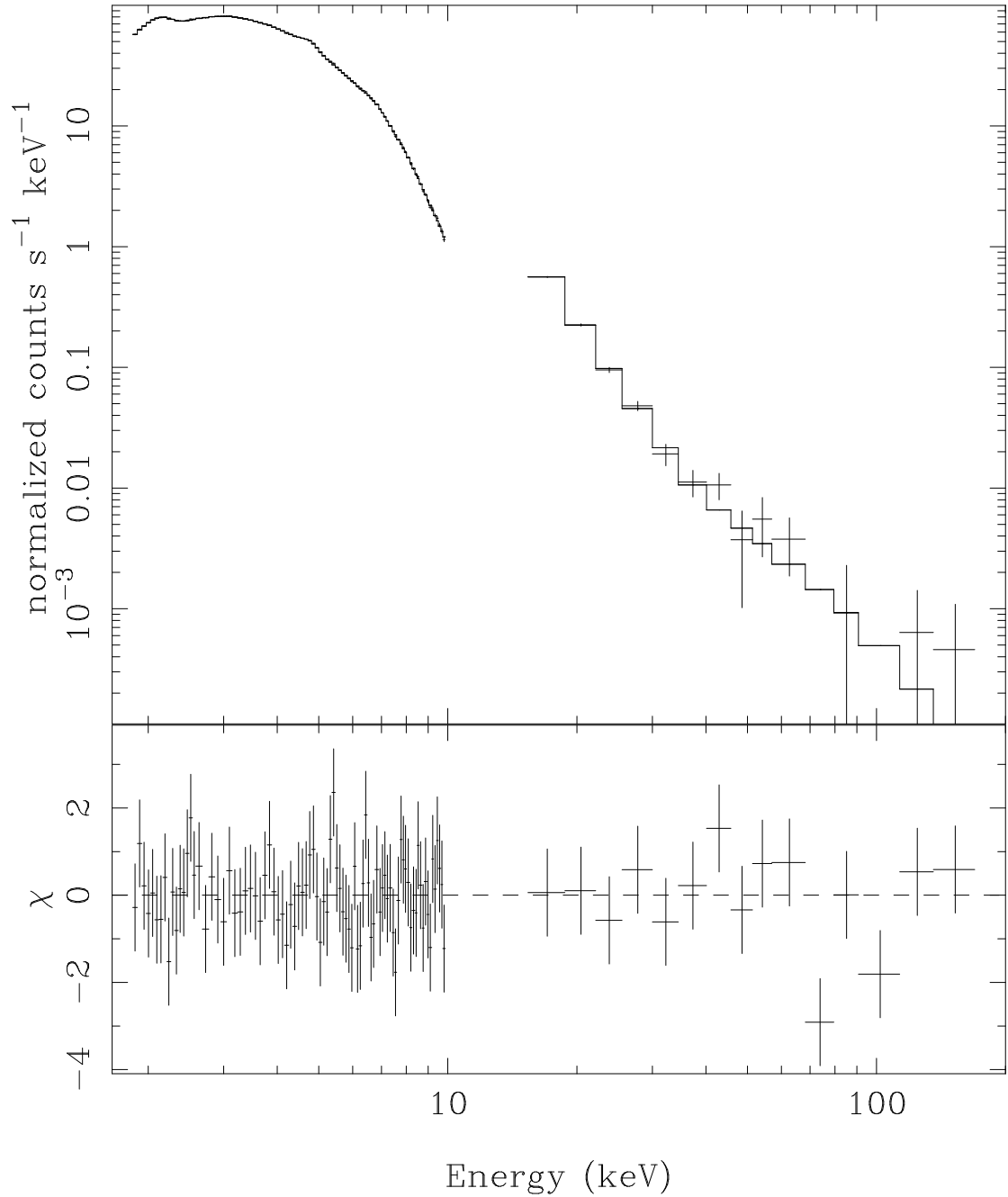}
\caption{\label{fig:fig5} Energy spectrum (1.8--200 keV) of Cir X--1 at the
orbital phase 0.61--0.63.  Data and the corresponding best fit model
(see Tab. 1, Model 3) are shown in the upper panel, residuals in units
of $\sigma$ with respect to the best fit model are shown in the lower
panel. }
\end{figure}

\begin{figure}
\plotone{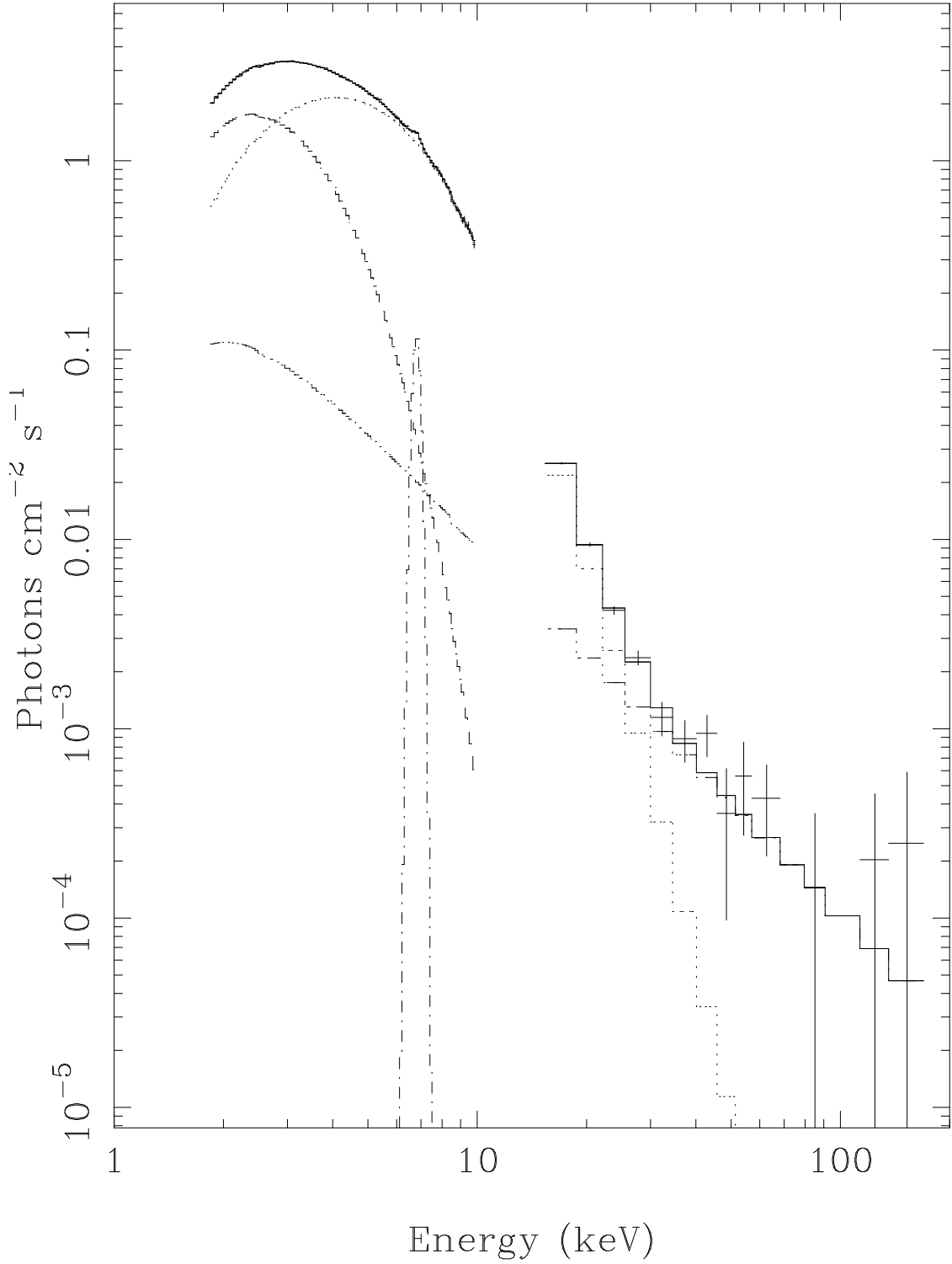}
\caption{\label{fig:fig6} Unfolded spectrum of Cir X--1 and the best fit model,
shown in this figure as a solid line. The single components of the
model are also shown, namely the blackbody (dashed line), {\it Comptt}
(dotted line), power-law (dot-dot-dashed line) and iron emission line
(dot-dashed line).  The absorption edge at 8.45 keV is visible. }
\end{figure}

\end{document}